# Dynamic Risk Assessment of Wildland-Urban Interface Fires

Yusheng Hu[1], Huaiyi Pan[1], Shaobo Zhong[2], Liying Zhang[3]

1. School of Management Science and Engineering, Beijing Information Science and Technology University, Beijing 102206, China;

2. Institute of Urban Systems Engineering, Beijing Academy of Science and Technology, Beijing 100089, China;

3. School of Science, China University of Mining and Technology-Beijing, Beijing 100083, China

**Abstract:** Wildland-Urban Interface (WUI) fires represent a compound disaster resulting from the interactions between natural ecosystems and human settlements, characterized by significantly dynamic evolving risks. However, most current risk assessment studies are based on static frameworks, which struggle to effectively capture the dynamic changes in risk over time. To address this issue, this paper proposes an innovative method that integrates a dynamic evaluation matrix, grey incidence analysis, and an optimization model for the dynamic risk assessment of WUI fires. This method incorporates time-series data by constructing a dynamic evaluation matrix, subsequently calculates the weighted standardized matrix for each evaluated area and its local volume matrices relative to the positive and negative ideal matrices. The dynamic differences between the evaluated areas and the ideal state are quantified by calculating the grey incidence degree, and an optimization model is established to solve for the superiority degree used for risk ranking. Research demonstrates that this method not only simplifies the computational process but also effectively captures the dynamic evolution patterns of fire risk across different areas, enabling refined risk classification. Compared to existing static methods, this framework overcomes their limitation in adequately representing risk dynamics, providing a more scientific basis for decision-making in the dynamic management and proactive prevention and control of WUI fires.

## 1. Introduction

The world currently faces three major crises: climate change, biodiversity loss, and environmental pollution, which necessitate a fundamental shift in how humanity coexists with nature. Given this context, incorporating urban forests into urban planning and construction has become imperative. This approach fosters harmony between urban areas and the natural environment, establishing a critical trajectory for ecological urban development in the new century. Indeed, enhancing urban forest initiative is the most crucial element of urban ecological projects, playing a pivotal role in improving urban ecology and guiding cities onto a civilized development path defined by sustainable production, improved livelihoods, and healthy ecosystems.

Data from the National Bureau of Statistics reveals that by the end of 2024, China's permanent resident urbanization rate stood at 67%. This figure has soared by 56% since the end of 1949, progressing at an average annual rate of 0.75%. This rapid urbanization has drastically escalated the demand for resources, infrastructure, and services, placing immense strain on natural ecosystems. In response to national development priorities and to advance the carbon peak and neutrality objectives, the construction of forest cities has emerged as an urgent imperative. A forest city is a healthy and stable ecosystem, predominantly of forests and trees, established within a city's administrative boundaries. It is characterized by urban-rural integration and compliance with standards for forest networks, forest health, ecological benefits, and culture, thereby achieving harmonious coexistence between humans and nature.

The development of the forest city concept was pioneered in North America, with the United States formalizing the term "forest city" in its 1962 Outdoor Recreation Resources Review. A decade later, the U.S. Urban Forestry Act established legally binding coverage benchmarks, requiring 27% urban forest coverage, 15% commercial area canopy cover, and 50% suburban forest coverage. This model was subsequently adopted and advanced throughout the developed world. China's engagement with this paradigm began in the 1990s. The movement was institutionalized in 2004 with the establishment of the "National Forest City" certification by the National Greening Committee and the State Forestry Administration (now the National Forestry and Grassland Administration). The parallel creation of the annual China Urban Forest Forum, guided by the principle of "Bring forests into cities and let cities embrace forests," served as a strategic catalyst. This framework aims to safeguard urban ecology, elevate city profiles, and underpin sustainable regional economic development. The outcomes have been substantial. The roster of National Forest Cities has expanded to 219 by the end of 2023. Nationally, forest coverage now stands at 24.02%, with China leading the globe in planted forest area. Urban areas have seen marked improvements in green coverage, park space per capita, and consecutive annual increases in forest stock, collectively enhancing biodiversity conservation efforts.

The construction of forest cities inevitably increases the extent of the Wildland-Urban Interface (WUI), a globally recognized zone of interface between human development and wildland vegetation. The concept was first defined in the 1970s by Stanford physicist C.P. Butler within forest fire research as the boundary where concentrations of natural fuels (trees, shrubs, grass) meet artificial fuels (buildings, structures) [1]. The US Federal Register later provided a formal definition, characterizing the WUI as the area where human development and activity meet or intermingle with undeveloped wildland vegetation [2]. This definition clarifies the WUI's three essential elements: human environments, wildland vegetation, and their dynamic interplay. WUI fires constitute a distinct category, markedly different from conventional forest fires or urban fires. The complex interplay of natural and anthropogenic factors in these areas creates an exceptionally high risk. Uncontrolled WUI fires frequently lead to disasters with substantial socio-economic losses.

Consequently, nations such as the United States, Canada, and Australia have long emphasized WUI fire research, a field that has demonstrably expanded in scale over recent decades. For China, the rapid advancement of its forest city initiative, alongside current limitations in fire prevention and suppression capacity, highlights a critical research gap. Undertaking systematic WUI fire risk assessments and formulating targeted control strategies is thus an imperative for improving governance efficacy and mitigating the frequency and consequences of these destructive fires.

Currently, most studies on WUI fire risk assessment focus on static evaluations, predominantly based on traditional risk assessment frameworks. These approaches typically conduct overlay analyses of fire hazards, vulnerability, and exposure levels but neglect the dynamic systemic nature of WUI fire risks. Given that these risks evolve over time, integrating dynamic assessment is crucial for enhancing the accuracy and scientific rigor of evaluations. To address this, our study employs the ideal matrix method to construct a similarity incidence analysis model for dynamic fire risk assessment. By calculating superiority degree to which the weighted standardized evaluation matrix of evaluated area aligns with the positive ideal matrix, this model ranks the fire risk of different areas and assigns corresponding risk levels. This method not only streamlines the computational process but also effectively captures the dynamic evolution of fire risks across various evaluated areas.

## 2. Literature Review

The risk assessment of WUI fire is fundamentally rooted in classic risk analysis and evaluation theories, representing an extension of the disaster risk assessment methodologies developed for forest fires and other natural hazards.

### 2.1. Definition of WUI

WUI represents a zone of convergence between wildland vegetation and human development, creating a unique constellation of fire hazards owing to the intermingling of combustible biomass and built structures. Globally, this risk is intensifying; urbanization and climate change are recognized drivers behind the increasing frequency and severity of WUI fires [2]. Establishing an operational definition of the WUI is therefore a foundational step for credible risk assessment. Pioneering methodologies for its demarcation utilize land-use and housing density data [3]. Subsequent advances in remote sensing and high-resolution data have significantly refined these techniques. A landmark contribution by Radeloff et al. (2018) synthesize satellite and census information to generate a global WUI map, thereby underscoring the magnitude of the challenge: 39% of new residential constructions in the U.S. are situated within these high-risk areas.

### 2.2. Fire Risk Assessment Methods

### 2.2.1. Traditional Statistical Models

Early research on fire risk assessment primarily relies on logistic regression and generalized linear models. For instance, Hanes et al. (2019) develop a coupled model using historical Canadian fire data, integrating climate, vegetation, and human activity factors. Their results identify the drought index and road density as pivotal predictive variables. In a more recent study, Liu et al. (2024) employ a time series prediction model enhanced with an attention mechanism to estimate fire occurrence probability. Furthermore, they integrate fire spread simulations using Cellular Automaton and Monte Carlo methods to assess potential burn areas. This integrated methodology facilitates a comprehensive risk assessment by evaluating both ignition likelihood and subsequent spread potential. By integrating multi-source data, such as optical/microwave remote sensing, meteorological, topographic, and human activity data, this method improves the model's sensitivity to crucial factors like vegetation water content.

Despite these advancements, such models remain limited in capturing complex nonlinear relationships and often depend on static datasets, which constrains their ability to represent dynamic fire risks effectively.

**2.2.2. Machine Learning**

In recent years, machine learning (ML) techniques such as random forest, support vector machine, and deep learning have been widely applied to enhance the accuracy of risk assessment. Xie et al. (2022) explore the potential of integrating multiple ML algorithms for building wildfire risk assessment models. They employ Bayesian optimization to tune the hyperparameters of several models, including random forest, support vector machine, and extreme gradient boosting. Moreover, they develop an integrated ML framework to generate wildfire risk grading maps and evaluate the importance of various triggering factors. Similarly, Tonbul (2024) introduces an innovative methodology for forest fire susceptibility mapping in the Mersin, Antalya, and Mugla provinces by combining ML models with explainable artificial intelligence. The study utilizes three advanced algorithms: extreme gradient boosting, gradient boosting machine, and light gradient boosting machine, to produce susceptibility maps based on 14 conditioning factors spanning meteorological, topographic, environmental, and anthropogenic variables. In a related effort, Naderpour et al. (2021) propose a spatial framework to quantify forest fire risk in the Northern Beaches region of Sydney. Their research identifies and spatially maps 36 significant indicators across multiple dimensions, including topography, morphology, climate, human activities, social factors, and physical characteristics. These indicators are used as inputs to an optimized deep neural network model, which improves the performance of the multilayer perceptron in assessing forest fire susceptibility.

Most machine learning methods are inherently static, relying on historical or cross-sectional data. They fail to effectively capture the dynamic evolution of risk factors (e.g., meteorology, vegetation water content) over time, thereby limiting their responsiveness to real-time changes in risk.

**2.2.3. Physics-based simulation model**

Physics-based fire behavior models, such as FARSITE and FlamMap, assess fire risk by simulating flame spread pathways and intensity. For example, Salis et al. (2021) integrate FARSITE with building vulnerability data to quantify potential losses in the WUI region of Sardinia, Italy. In a study focused on wood-frame contiguous villages in Western Hunan, Zhang et al. (2022) develop a comprehensive risk assessment framework that accounts for multiple factors, including village layout, street configuration, building materials, structural forms, energy use patterns, and fire-related behaviors. The researchers employ fire dynamics simulator (FDS) and SketchUp to visualize and analyze fire combustion dynamics at both single-building and cluster scales, providing a basis for targeted mitigation strategies. Simulation results derived from this approach can support the optimization of fire protection measures for such vulnerable settlements.

While physics-based simulation models can mimic dynamic processes, their high computational costs and stringent requirements for input data accuracy make them difficult to apply to large-scale, rapid dynamic risk assessments.

**2.2.4. Fire risk assessment model based on information technology**

Recent years have witnessed the continued emergence of innovative fire risk assessment methods leveraging information technology, applied in both built environments and natural ecosystems. Through technological integration and interdisciplinary collaboration, significant improvements have been made in fire prevention and control capabilities. For instance, Wang et al. (2021) apply the fire risk assessment method for engineering (FRAME) to develop an evaluation index system for buildings during their operation and maintenance phases. This system considers potential risk levels, acceptable risk thresholds, and protection levels. Furthermore, the authors establish a computational model for quantifying fire risk in these phases by incorporating building information modeling (BIM) technology, and define a risk valuation standard to enhance building fire resilience. Their results demonstrate that BIM can be effectively utilized in fire safety management, providing a more intuitive and scientific basis for fire risk prevention and control. In a natural ecosystem context, Nuthammachot and Stratoulias (2021) develop a GIS-based multi-criteria decision analysis system for fire risk assessment and fire potential mapping in a peat swamp forest in Hua Sai District, Thailand. The study analyzes 55 recorded fire occurrences in peat swamp areas between 2012 and 2016. Using the analytic hierarchy process (AHP) integrated with GIS, the authors evaluate a range of contributing factors, including elevation, slope, aspect, precipitation, proximity to rivers and settlements, and land use patterns. The findings reveal that the fire risk zones identified by the proposed method align closely with historical fire incident data.

**2.2.5. Fuzzy mathematical model**

Fuzzy mathematics has emerged as a pivotal methodology for risk modeling in complex fire scenarios, effectively quantifying the uncertainty, fuzziness, and multi-factor coupling characteristics inherent in fire risk systems. Huang et al. (2022) introduce a novel fire risk assessment approach for lithium-ion batteries during transportation and storage. Their method identifies potential failure paths and basic events, and evaluates each of them using likelihood, severity, and hazard control number to derive a hazard risk number (HRN) for fire accidents. These indicators are further broken down into sub-indicators, which are synthesized using a fuzzy logic framework to ensure comprehensive evaluation. To enhance battery safety, threshold limits are proposed for four aspects: the synthesized HRN, HRNs of all failure paths, and the likelihood and severity of all basic events. A case study on marine battery transportation confirms that the method accurately assesses fire risks and pinpoints potential issues.

In a related development, Wang et al. (2024) propose an innovative fire risk assessment model for large-scale commercial and high-rise buildings using an intuitionistic fuzzy set approach integrated with social graph theory. Their method provides a scientific and objective weighting mechanism that synthesizes positive, negative, and hesitant judgments from multiple experts while accounting for their individual importance. To validate the model, fire safety data is collected from 11 buildings. Comparative analysis with existing methods demonstrates the robustness and reliability of the proposed weighting strategy.

Information technology-based models and fuzzy mathematics, although excellent in handling spatial information and uncertainty, often produce assessment results that are static "snapshots," lacking continuous characterization of risk sequences.

**2.2.6. Multi-criteria decision making model**

In the field of fire risk assessment, multi-criteria decision-making (MCDM) models have become essential tools for balancing technical feasibility, economic cost, and social benefits in complex scenarios, owing to their capacity to integrate multidimensional indicators and dynamically assign weights. In recent years, research has increasingly incorporated heterogeneous factors, such as hazard drivers, exposure, vulnerability, and emergency response capacity, using methods like the AHP and the Technique for order preference by similarity to ideal solution (TOPSIS). These developments have enabled a shift in risk assessment from single-threshold judgments toward multi-objective collaborative optimization. For example, Gulum et al. (2021) conceptualize post-earthquake fire risk assessment as an MCDM problem and propose a two-level evaluation framework. The authors first determine the importance weights of relevant criteria using the interval-valued neutrosophic AHP (IVN-AHP) method. They then apply the interval-valued neutrosophic TOPSIS (IVN-TOPSIS) approach to rank districts in the Anatolian side of Istanbul according to their post-earthquake fire risk levels. The proposed methodology is implemented with real-world data to identify the highest-risk districts in Istanbul, Turkey.

In another study, Parvar et al. (2024) employ a multi-criteria evaluation method to generate a fire risk map by analyzing both natural and anthropogenic factors, enabling the identification of vulnerable areas. The study further examines the relationship between fire occurrences and key influencing element, such as meteorological conditions, land surface temperature (LST), and precipitation, with the occurrence of fire in different years. Using CHIRPS and Landsat data, LST changes and precipitation are assessed, while MODIS products are utilized to track 23-year variations in fire-affected areas within the study region. Although multi-criteria decision-making models can effectively integrate multidimensional indicators, their traditional forms often lack consideration for the performance differences of evaluation objects across time series during weight setting and alternative ranking. This makes it difficult to handle dynamic evaluation matrices and thus fails to reflect the fluctuation of risk levels over time.

Existing methods exhibit significant shortcomings in achieving an assessment framework that can both comprehensively consider multi-source heterogeneous indicators and effectively quantify the temporal dynamics of risk. To bridge this research gap, this paper proposes an innovative method that integrates dynamic evaluation matrices, grey incidence analysis, and an optimization model. This method incorporates time-series data by constructing a dynamic evaluation matrix, utilizes local volume matrices and grey incidence degree to quantify the dynamic differences between the evaluated areas and the ideal state, and finally solves for the superiority degree through an optimization model. This enables the dynamic and refined ranking and classification of WUI fire risks. This framework aims to overcome the static limitations of existing methods and provide a more scientific basis for decision-making in the dynamic management and proactive prevention and control of fire risk.

## 3. Comprehensive Analysis of Critical Risk Factors for WUI Fires

WUI fires constitute a compound disaster paradigm, originating from the complex interactions at the nexus of natural ecosystems and human settlements. In the context of ongoing global urbanization and intensifying climate change, WUI fire risks are undergoing a marked non-linear intensification. The occurrence of these fires is attributable to a constellation of factors, including vegetation fuel loads, structural attributes, anthropogenic ignition sources, topographic features, and climatic variables, that act in concert to drive fire behavior.

### 3.1. Risk Factors Related to Vegetation

The spatial continuity of vegetation plays a critical role in facilitating fire spread by forming continuous "fuel corridors." In a WUI fire risk assessment, Ghermandi (2016) emphasizes the influence of vegetation type (e.g., forest and grassland), fuel load, and spatial continuity on fire propagation. Specifically, forests with low horizontal vegetation continuity exhibit a reduced likelihood of fire spread. Syphard et al. (2017) apply the Integral Index of Connectivity to WUI areas in the western United States and find that for every 10% increase in continuous vegetation cover, the rate of fire spread increases by 22%. Notably, in forest-grassland transition zones, highly continuous vegetation can enable fires to cross natural barriers such as rivers and bare ground, thereby directly threatening human settlements. Furthermore, Syphard et al. (2024) explore geographical variations in fire regimes under climate change, identifying terrain, climate, and vegetation distribution as key influencing factors.

Variations in the physical and chemical properties of different vegetation types significantly affect ignition thresholds and fire intensity. For example, coniferous forests such as pine and eucalyptus exhibit fire intensities 2-3 times greater than those in broad-leaved forests due to their high volatile oil content (Keeley & Syphard, 2019). In arid environments, shrubs such as California sagebrush function as "flammable barriers" at the WUI due to their extremely low moisture content (<5%) and dense canopy structure (Fernandes et al., 2013). Syphard et al. (2022) investigate the transition from shrubland to grassland in Southern California, identifying short-interval fires as the primary driver and evaluating the implications of such vegetation shifts on fire risk.

## 3.2. Risk Factors Associated with Buildings and Human Activities

Low-density scattered development (1-5 buildings per hectare) significantly increases fire exposure risk. Radeloff et al. (2018) use GIS simulations to demonstrate that dispersed building layouts reduce the effectiveness of firebreaks by 60% and decrease the efficiency of firefighting resource allocation by 35%. Moreover, combustible building materials such as wooden roofs and fences can exacerbate local fire outbreaks when intermingled with vegetation (Cohen, 2010).

Road network design critically affects emergency response capabilities. Calkin et al. (2014) emphasize that roads narrower than 4 meters or gated community entrances in WUI areas can delay the arrival of firefighting vehicles by 30-50%. Aging power infrastructure also poses a significant hazard; Balch et al. (2017) report that 12-25% of WUI fires in the United States are caused by power line faults or transformer failures.

Human-caused ignition sources, including open-air burning and vehicle sparks, account for over 80% of WUI fire ignitions (Syphard et al., 2017). Calviño et al. (2017) investigate the interacting factors influencing fire ignition risk, including vegetation type, topographic characteristics, and WUI location. Additionally, they examine human activities and motivations associated with fires, assessing whether these factors vary by vegetation type and WUI context. Their results reveal significant interactions among topography, vegetation type, and WUI location.

## 3.3. Risk Factors of Topography and Climate

Topography and climate significantly influence the ignition and spread of WUI fires. Slope is a key factor governing the speed and intensity of fire propagation. Steep slopes substantially accelerate upslope flame spread by altering heat radiation pathways and airflow patterns. Alexandre et al. (2016), through field observations and numerical simulations, demonstrate that for every 10-degree increase in slope, the rate of fire spread increases by 15-20%, while fire line intensity can rise by more than 30%. In valley terrain, the combined effect of slope and canyon-induced airflow creates a "chimney effect," dramatically increasing the vertical velocity of flames and leading to high-intensity, uncontrollable wildfires.

Aspect indirectly affects vegetation flammability by regulating solar radiation exposure. Moritz et al. (2014), analyzing temperate forest fire data across the Northern Hemisphere, find that sun-facing slopes (e.g., south-facing) experience prolonged solar exposure, leading to lower surface evaporation and vegetation moisture content that is 20-40% lower than on shaded slopes, thereby increasing flammability. Additionally, sun-facing slopes often support coniferous forests with low moisture and high volatile content, further elevating fire probability and spread rates.

Long-term drought and extreme weather events serve as critical drivers of WUI fires. During the severe drought in California from 2012 to 2016, the Climatic Water Deficit (CWD) increased by 50%, resulting in a 2.3-fold increase in fire frequency compared to the historical average (Williams et al., 2019). Strong winds exceeding 30 km/h can propagate fires through ember transport, extending the fire front distance to 4-6 times that under calm wind conditions (Calkin et al., 2014). Jolly et al. (2015) further show that the global Fire Weather Index (FWI) increased by 35% between 1979 and 2013, significantly prolonging the high-risk period for WUI fires. Moreover, Abatzoglou et al. (2019) indicate that rising temperatures and shifting precipitation patterns not only lengthen the fire season but also expand the potential risk area for WUI fires.

## 4. Fire Risk Assessment Model Based on Volumetric Incidence Degree
### 4.1. Principles and Methods of Fire Risk Assessment Model

Assuming that the dynamic fire risk assessment for WUI involves $n$ evaluated areas $U_1, U_2, \ldots, U_n$ and $m$ evaluation indices $E_1, E_2, \ldots, E_m$ with corresponding weights $\lambda_1, \lambda_2, \ldots, \lambda_m$. Additionally, the time weights of each evaluated area over $T$ periods are $\theta_1, \theta_2, \ldots, \theta_T$. Let the evaluation value of the $i$-th evaluated area $U_i$ for index $E_j$ in period $t$ be denoted as $A_{ij}(t)$ $(i=1,2,\ldots,n;\ j=1,2,\ldots,m;\ t=1,2,\ldots,T)$. Then, the evaluation matrix of all evaluated areas under risk evaluation indices $E_1, E_2, \ldots, E_m$ in period $t$ can be expressed as

$$A(t) = \begin{bmatrix} a_{11}(t) & q_{12}(t) & \cdots & a_{1m}(t) \\ a_{21}(t) & a_{22}(t) & \cdots & a_{2m}(t) \\ \cdots & \cdots & \cdots & \cdots \\ a_{n1}(t) & a_{n2}(t) & \cdots & a_{nm}(t) \end{bmatrix}$$

Here, the $i$-th row represents the evaluation values of $U_i$ for indices $E_1, E_2, \ldots, E_m$ in period $t$. The evaluation values of $U_i$ across all indices over $T$ periods can be represented as a matrix:

$$A_i = (A_{i1}, A_{i2}, \ldots, A_{iT}) = \begin{bmatrix} a_{i1}(1) & a_{i1}(2) & \cdots & a_{i1}(T) \\ a_{i2}(1) & a_{i2}(2) & \cdots & a_{i2}(T) \\ \cdots & \cdots & \cdots & \cdots \\ a_{im}(1) & a_{im}(2) & \cdots & a_{im}(T) \end{bmatrix} \quad (1)$$

This is referred to as the dynamic evaluation matrix of $U_i$.

Due to potential differences in measurement units and orders of magnitude among the evaluation indices, direct dynamic risk assessment of WUI is not feasible. Therefore, it is necessary to standardize the dynamic evaluation matrices $A_i\ (i=1,2,\ ...,n)$ for each evaluated area. If the evaluation index $E_j$ is benefit-oriented (i.e., a larger value indicates higher risk), then $a_{ij}(t)$ can be standardized as

$$b_{ij}(t) = \frac{a_{ij}(t) - \min\limits_{i}\min\limits_{t} a_{ij}(t)}{\max\limits_{i}\max\limits_{t} a_{ij}(t) - \min\limits_{i}\min\limits_{t} a_{ij}(t)} \tag{2}$$

If the evaluation index $E_j$ is cost-oriented (i.e., a larger value indicates lower risk), then $a_{ij}(t)$ can be standardized as

$$b_{ij}(t) = \frac{\max\limits_{i}\max\limits_{t} a_{ij}(t) - a_{ij}(t)}{\max\limits_{i}\max\limits_{t} a_{ij}(t) - \min\limits_{i}\min\limits_{t} a_{ij}(t)} \tag{3}$$

If the evaluation index $E_j$ is intermediate-type (i.e., value closer to the median indicates higher risk), then $a_{ij}(t)$ can be standardized as

$$b_{ij}(t) = 1 - \frac{|a_{ij}(t) - M_j(t)|}{\max\limits_{i}\max\limits_{t}|a_{ij}(t) - M_j(t)|} \tag{4}$$

where $M_j$ represents the median value of the intermediate-type risk index $E_j$ during period $t$. If the evaluation index $E_j$ is interval-type (i.e., a larger value within a specific interval indicates higher risk), then $a_{ij}(t)$ can be standardized to

$$b_{ij}(t) = \begin{cases} 1 - \dfrac{a_{ij}^l(t) - a_{ij}(t)}{\max\{a_{ij}^l(t) - \min\limits_{i}\min\limits_{t} a_{ij}(t), \max\limits_{i}\max\limits_{t} a_{ij}(t) - a_{ij}^u(t)\}}, & a_{ij}(t) < a_{ij}^l(t) \\ 1 & a_{ij}^l(t) \leq a_{ij}(t) \leq a_{ij}^u(t) \\ 1 - \dfrac{a_{ij}(t) - a_{ij}^u(t)}{\max\{a_{ij}^l(t) - \min\limits_{i}\min\limits_{t} a_{ij}(t), \max\limits_{i}\max\limits_{t} a_{ij}(t) - a_{ij}^u(t)\}}, & a_{ij}^u(t) < a_{ij}(t) \end{cases}$$

After standardization, the standardized evaluation matrix for indices $E_1, E_2, ..., E_m$ in period $t$ is obtained for each evaluated area:

$$B(t) = \begin{bmatrix} b_{11}(t) & b_{12}(t) & \cdots & b_{1m}(t) \\ b_{21}(t) & b_{22}(t) & \cdots & b_{2m}(t) \\ \cdots & \cdots & \cdots & \cdots \\ b_{n1}(t) & b_{n2}(t) & \cdots & b_{nm}(t) \end{bmatrix}$$

In this matrix, the *i*-th row represents the standardized evaluation values of area $U_i$ for indices $E_1, E_2, ..., E_m$ in period $t$.

The standardized evaluation values of area $U_i(i=1,2,...,n)$ under $m$ evaluation indices across $T$ periods are represented as a standardized dynamic evaluation matrix:

$$B_i = (B_{i1}, B_{i2},...,B_{iT}) = \begin{bmatrix} b_{i1}(1) & b_{i1}(2) & \cdots & b_{i1}(T) \\ b_{i2}(1) & b_{i2}(2) & \cdots & b_{i2}(T) \\ \cdots & \cdots & \cdots & \cdots \\ b_{im}(1) & b_{im}(2) & \cdots & b_{im}(T) \end{bmatrix}.$$

The index weight vector $\Lambda = (\lambda_1, \lambda_2,...,\lambda_m)$ and the time weight vector $\Theta = (\theta_1, \theta_2,...,\theta_T)$ are multiplied by the standardized dynamic evaluation matrix of area $U_i(i=1,2,...,n)$ to obtain the weighted standardized evaluation matrix:

$$C_i = (c_{ij}(t))_{m \times T} = (\lambda_j b_{ij}(t)\theta_t)_{m \times T} = \begin{bmatrix} \lambda_1 b_{i1}(1)\theta_1 & \lambda_1 b_{i1}(2)\theta_2 & \cdots & \lambda_1 b_{i1}(T)\theta_T \\ \lambda_2 b_{i2}(1)\theta_1 & \lambda_2 b_{i2}(2)\theta_2 & \cdots & \lambda_2 b_{i2}(T)\theta_T \\ \cdots & \cdots & \cdots & \cdots \\ \lambda_m b_{im}(1)\theta_1 & \lambda_m b_{im}(2)\theta_2 & \cdots & \lambda_m b_{im}(T)\theta_T \end{bmatrix}. \quad (5)$$

This study employs the weighted standardized evaluation matrix and a similarity matrix incidence degree model to conduct dynamic fire risk assessments. Let $c_j^+(t)$ and $c_j^-(t)$ denote the maximum and minimum evaluated values of index $E_j$ for all evaluated areas across different periods, where

$$c_j^+(t) = \max_i c_{ij}(t),\ c_j^-(t) = \min_i c_{ij}(t),\ j=1,2,...,m; t=1,2,...,T \quad (6)$$

Using $c_j^+(t)$ and $c_j^-(t)$ as fundamental elements, the weighted matrices for dynamic fire risk assessment are constructed as:

$$C^+ = \begin{bmatrix} c_1^+(1) & c_1^+(2) & \cdots & c_1^+(T) \\ c_2^+(1) & c_2^+(2) & \cdots & c_2^+(T) \\ \cdots & \cdots & \cdots & \cdots \\ c_m^+(1) & c_2^+(m) & \cdots & c_m^+(T) \end{bmatrix} \text{ and } C^- = \begin{bmatrix} c_1^-(1) & c_1^-(2) & \cdots & c_1^-(T) \\ c_2^-(1) & c_2^-(2) & \cdots & c_2^-(T) \\ \cdots & \cdots & \cdots & \cdots \\ c_m^-(1) & c_2^-(m) & \cdots & c_m^-(T) \end{bmatrix}$$

Here, $C^+$ and $C^-$ represent the positive ideal matrix and the negative ideal matrix, respectively, constructed from the maximum and minimum values of all evaluation indices for all areas in each period. In dynamic fire risk assessment, the evaluated area corresponding to the positive ideal matrix represents the highest risk scenario. However, it is rare for the weighted matrix of an evaluated area to perfectly align with either the positive or negative ideal matrices. Therefore, the risk level of each evaluated area can be determined by evaluating its proximity to these ideal matrices. This study applies a similarity incidence analysis model to address the issue of risk ranking for each evaluated area in the dynamic fire risk assessment.

## 4.2. Incidence Degree Between the Dynamic Evaluation Matrix of Each Evaluated Area and the Positive-Negative Ideal Matrix

The absolute incidence degree of a matrix, derived from sequence incidence degree, does not fully satisfy the completeness axioms of incidence or the order-preserving property under scalar multiplication transformations. To overcome these limitations, this study adopts the volumetric incidence degree to measure the closeness between the dynamic evaluation matrix of the WUI and the positive/negative ideal matrices.

**Definition 1** Let $C = (c_{ij})_{m \times n}$ denote a behavior matrix, and $\tilde{C} = (\tilde{c}_{ij})_{m \times n}$ represent the zero-starting-point image of $C$. Define

$$d_{ij} = \frac{\tilde{c}_{ij} + \tilde{c}_{i+1,j+1}}{6} + \frac{\tilde{c}_{i+1,j} + \tilde{c}_{i,j+1}}{3} \tag{7}$$

Then $D = (d_{ij})_{(m-1) \times (n-1)}$ is referred to as the local volume matrix of $B$.

**Definition 2** Assume that the local volume matrices of the behavior matrix of the reference factor, $C^{(0)} = (c_{ij}^{(0)})_{m \times n}$, and the behavior matrix of factor $k$, $C^{(k)} = (c_{ij}^{(k)})_{m \times n}$ $(k = 1, 2, ..., s)$, are given by $D^{(k)} = (d_{ij}^{(k)})_{(m-1) \times (n-1)}$. Let

$$d_{ij}^{0k} = |d_{ij}^{(0)} - d_{ij}^{(k)}| \tag{8}$$

Then the matrix $D^{0k} = (d_{ij}^{0k})_{(m-1) \times (n-1)}$ is referred to as the volume difference matrix between factor $k$ and the reference factor over the corresponding region. Let

$$d_{\max} = \max_i \max_j \max_k d_{ij}^{0k}, \quad d_{\min} = \min_i \min_j \min_k d_{ij}^{0k}.$$

Then $d_{\max}$ and $d_{\min}$ are called the maximum volume difference and the minimum volume difference, respectively.

The volume difference matrix serves as a tool to measure the similarity between two factors within a specific region. A larger volume difference indicates a greater disparity in the changing trends of the two factors in that region.

**Definition 3** Let $C^{(0)} = (c_{ij}^{(0)})_{m \times n}$ and $C^{(k)} = (c_{ij}^{(k)})_{m \times n}$ $(k = 1, 2, ..., s)$ be the behavioral matrix of the reference factor and factor $k$, respectively. The volume difference matrix between $C^{(0)}$ and $C^{(k)}$ is denoted as $D^{0k} = (d_{ij}^{0k})_{(m-1) \times (n-1)}$, with maximum and minimum elements being $d_{\max}$ and $d_{\min}$, respectively. The grey incidence coefficient matrix $G^{(k)} = (g_{ij}^{(k)})_{(m-1) \times (n-1)}$ of factor $k$ relative to the reference factor is then defined by

$$g_{ij}^{(k)} = \begin{cases} 1, & d_{max} = 0 \\ \dfrac{d_{max} - d_{ij}^{0k}}{d_{max} - d_{min}}, & d_{max} \neq 0 \end{cases} \quad (9)$$

The volumetric incidence degree of factor $k$ with respect to the reference factor is given by

$$\gamma(C^{(0)}, C^{(k)}) = \dfrac{\sum_{i=1}^{m} \sum_{j=1}^{n} g_{ij}^{(k)}}{(m-1)(n-1)} \quad (10)$$

Since the volumetric incidence degree incorporates the influence of known factors on the incidence degree, it satisfies the wholeness axiom of incidence. Moreover, as this incidence model assumes local similarity among matrices, it effectively avoids misjudgment in the incidence degree for fluctuating behavior matrices.

### 4.3. Optimizing Dynamic Fire Risk Assessment for the WUI

Let $\gamma_i^+$ denote the similarity incidence degree between the weighted standardized evaluation matrix of the evaluated area $U_i$ and the positive ideal matrix, and $\gamma_i^-$ ($i = 1, 2, ..., n$) represent that with the negative ideal matrix. A larger value of $\gamma_i^+$ indicates that the weighted standardized evaluation matrix of $U_i$ is closer to the positive ideal matrix $C^+$, implying a higher risk level for the area. Conversely, a larger value of $\gamma_i^-$ suggests that the matrix is closer to the negative ideal matrix $C^-$, indicating a lower risk level. Typically, the evaluated area with the highest risk exhibits the strongest similarity to the positive ideal matrix and the weakest similarity to the negative ideal matrix. Let $s_i$ denote the superiority degree to which the weighted standardized evaluation matrix of $U_i$ belonging to the positive ideal matrix $C^+$. Consequently, the superiority degree of the same matrix belonging to the negative ideal matrix $C^-$ is $1 - s_i$. To determine the superiority degree, the following optimization model is established:

$$\min H(\mathbf{s}) = \sum_{i=1}^{n} \{[(1-s_i)\gamma_i^+]^2 + (s_i \gamma_i^-)^2\},$$

where $\mathbf{s} = (s_1, s_2, ..., s_n)$ is the vector of superiority degrees for all evaluated areas. Let

$$\dfrac{\partial H(\mathbf{s})}{\partial s_i} = 0.$$

Then, it follows that

$$\dfrac{\partial H(\mathbf{s})}{\partial s_i} = -2(1-s_i)(\gamma_i^+)^2 + 2s_i(\gamma_i^-)^2.$$

Consequently, $s_i = \dfrac{(\gamma_i^+)^2}{(\gamma_i^+)^2 + (\gamma_i^-)^2}$. (11)

Since $s_i$ reflects the degree to which the weighted standardized evaluation matrix of evaluated area $U_i$ aligns with the positive ideal matrix, a larger value of $s_i$ indicates a higher risk level for $U_i$. Therefore, the superiority degree can be used as a criterion for ranking fire risks in the WUI.

### 4.4. Determination of the Comment Set

In the fire dynamic risk assessment, the risk levels are categorized into 7 grades: {extremely low risk, low risk, slightly low risk, medium risk, slightly high risk, high risk, extremely high risk}. The corresponding risk comment set $V$ for these evaluated areas is defined as:

$$V = \{v_1, v_2, v_3, v_4, v_5, v_6, v_7\} = \{0.1, 0.2, 0.4, 0.6, 0.7, 0.8, 0.9\}.$$

The risk degree in the comment set is directly proportional to the risk level. If a risk degree falls between two adjacent risk level, the higher level is adopted. For instance, if the risk degree is 0.3, the risk level is classified as "slightly low risk".

Based on the above analysis, the fire risk assessment process can be summarized in the following steps:

(1) Determine the dynamic evaluation matrix $A_i$ for each evaluated area.

(2) Normalize the dynamic evaluation matrix $A_i$ to obtain the standardized dynamic evaluation matrix $B_i$.

(3) Calculate the weighted standardized evaluation matrix $C_i$ for each evaluated area, as well as the positive and negative ideal matrices $C^+$ and $C^-$.

(4) Calculate the local volume matrix $D_i$ for each weighted standardized evaluation matrix, and the local volume matrices $D^+$ and $D^-$ for the positive and negative ideal matrices.

(5) Determine the volume difference matrices $D_{(i)}^+$ and $D_{(i)}^-$ between the local volume matrix of each evaluated area and the local volume matrices of the positive and negative ideal matrices.

(6) Calculate the grey incidence coefficient matrices $G_{(i)}^+$ and $G_{(i)}^-$ for each volume difference matrix $D_{(i)}^+$ and $D_{(i)}^-$.

(7) Calculate the incidence degrees $\gamma_i^+$ and $\gamma_i^-$ for grey incidence coefficient matrices $G_{(i)}^+$ and $G_{(i)}^-$, respectively.

(8) Establish an optimization model for WUI fire risk assessment to derive the superiority degree $s_i$ for each evaluated area.

(9) Compare the superiority degrees $s_i$ of each evaluated area to determine the risk ranking and the risk levels.

## 5. Case Study Analysis

This section presents dynamic risk assessments at 6 different time points for 3 evaluated areas as case studies. The study involves an extensive review of relevant domestic and international literature, supplemented by in-depth surveys and interviews. Based on the findings, an evaluation index system for assessing the risk of fire occurrence in the WUI is developed, as shown in the table below.

**Table 1: Index System for Assessing Fire Occurrence Risk in WUI**

| Primary Index | Secondary Index |
|---|---|
| the Condition of Combustible Materials in the WUI | the Fuel Load in the WUI |
| | Moisture Content of Combustible Materials in the WUI |
| | Spatial Distribution of Combustible Materials in the WUI |
| Surrounding Environment of the Transitional Zone in the WUI | Uncontrolled Fire Spread in Agricultural, Forestry, and Livestock Production Regions |
| | Domestic Fire Use in Daily Life |
| | Population Density Distribution |
| | Road Network Density |
| Meteorological Conditions in the WUI | Precipitation Levels |
| | Relative Humidity |
| | Air Temperature |
| | Wind Velocity |
| Topographical Characteristics in the WUI | Slope Gradient |
| | Slope Aspect |
| | Topographic Position |
| | Elevation Above Sea Level |

Using the scaling expansion method, the weight vector of the index system can be determined as

$\omega$=(0.1458,0.1303,0.1114,0.0666,0.0612,0.0585,0.0559,0.0650,0.0542,0.0451, 0.0376,0.0450,0.0430,0.0411,0.0392)

The weight vector corresponding to the 6 time points is given by

$$\mu = (0.21, 0.15, 0.25, 0.12, 0.18, 0.09)$$

Based on expert ratings for the 15 indices, the dynamic assessment matrices for the 3 evaluated areas across 6 time points are denoted as

$$A_1 = \begin{bmatrix} 20 & 30 & 35 & 40 & 25 & 50 & 45 & 20 & 55 & 40 & 60 & 70 & 65 & 80 & 60 \\ 30 & 40 & 45 & 45 & 30 & 55 & 50 & 30 & 60 & 50 & 55 & 70 & 65 & 80 & 60 \\ 25 & 35 & 30 & 35 & 40 & 50 & 55 & 35 & 50 & 45 & 65 & 70 & 65 & 80 & 60 \\ 30 & 40 & 50 & 40 & 35 & 45 & 35 & 25 & 60 & 55 & 50 & 70 & 65 & 80 & 60 \\ 40 & 30 & 45 & 50 & 30 & 55 & 40 & 45 & 65 & 40 & 50 & 70 & 65 & 80 & 60 \\ 45 & 50 & 35 & 30 & 45 & 40 & 50 & 35 & 45 & 35 & 65 & 70 & 65 & 80 & 60 \end{bmatrix}^T,$$

$$A_2 = \begin{bmatrix} 25 & 35 & 40 & 35 & 30 & 45 & 40 & 25 & 50 & 45 & 55 & 60 & 75 & 70 & 65 \\ 35 & 30 & 35 & 40 & 35 & 50 & 45 & 35 & 55 & 40 & 65 & 60 & 75 & 70 & 65 \\ 30 & 40 & 25 & 45 & 40 & 55 & 50 & 30 & 45 & 50 & 55 & 60 & 75 & 70 & 65 \\ 35 & 45 & 30 & 35 & 30 & 40 & 30 & 35 & 40 & 30 & 50 & 60 & 75 & 70 & 65 \\ 45 & 35 & 35 & 30 & 35 & 50 & 45 & 45 & 60 & 45 & 40 & 60 & 75 & 70 & 65 \\ 40 & 45 & 30 & 40 & 45 & 40 & 35 & 30 & 40 & 50 & 55 & 60 & 75 & 70 & 65 \end{bmatrix}^T$$

$$A_3 = \begin{bmatrix} 40 & 50 & 30 & 20 & 65 & 50 & 55 & 30 & 40 & 55 & 50 & 80 & 70 & 60 & 50 \\ 30 & 40 & 25 & 30 & 60 & 55 & 50 & 40 & 50 & 50 & 60 & 80 & 70 & 60 & 50 \\ 35 & 45 & 35 & 40 & 55 & 60 & 60 & 45 & 55 & 40 & 65 & 80 & 70 & 60 & 50 \\ 45 & 35 & 40 & 45 & 50 & 45 & 65 & 50 & 45 & 35 & 45 & 80 & 70 & 60 & 50 \\ 50 & 55 & 45 & 35 & 45 & 55 & 55 & 55 & 65 & 40 & 45 & 80 & 70 & 60 & 50 \\ 45 & 40 & 35 & 25 & 40 & 65 & 40 & 35 & 45 & 45 & 50 & 80 & 70 & 60 & 50 \end{bmatrix}^T$$

Here, a higher score for an index indicates a higher level of risk in the evaluated area.

From expression (2), the standard dynamic evaluation matrix can be obtained by normalizing the dynamic evaluation matrix. The resulting standardized dynamic evaluation matrices are as follows:

$$B_1 = \begin{bmatrix} 0 & 0 & 0.4 & 0.67 & 0 & 0.4 & 0.43 & 0 & 0.6 & 0.4 & 0.8 & 0.88 & 0 & 1 & 0.67 \\ 0.33 & 0.4 & 0.8 & 0.83 & 0.13 & 0.6 & 0.57 & 0.29 & 0.8 & 0.8 & 0.6 & 0.88 & 0 & 1 & 0.67 \\ 0.17 & 0.2 & 0.2 & 0.5 & 0.38 & 0.4 & 0.71 & 0.43 & 0.4 & 0.6 & 1 & 0.88 & 0 & 1 & 0.67 \\ 0.33 & 0.4 & 1 & 0.67 & 0.25 & 0.2 & 0.14 & 0.14 & 0.8 & 1 & 0.4 & 0.88 & 0 & 1 & 0.67 \\ 0.67 & 0 & 0.8 & 1 & 0.13 & 0.6 & 0.29 & 0.71 & 1 & 0.4 & 0.4 & 0.88 & 0 & 1 & 0.67 \\ 0.5 & 0.36 & 0.2 & 0.2 & 0.31 & 0 & 0.31 & 0.27 & 0.08 & 0.9 & 0.38 & 0.88 & 0 & 1 & 0.67 \end{bmatrix}^T$$

$$B_2 = \begin{bmatrix} 0.17 & 0.2 & 0.6 & 0.5 & 0.13 & 0.2 & 0.29 & 0.14 & 0.4 & 0.6 & 0.6 & 0.75 & 1 & 0.5 & 1 \\ 0.5 & 0 & 0.4 & 0.67 & 0.25 & 0.4 & 0.43 & 0.43 & 0.6 & 0.4 & 1 & 0.75 & 1 & 0.5 & 1 \\ 0.33 & 0.4 & 0 & 0.83 & 0.38 & 0.6 & 0.57 & 0.29 & 0.2 & 0.8 & 0.6 & 0.75 & 1 & 0.5 & 1 \\ 0.5 & 0.6 & 0.2 & 0.5 & 0.13 & 0 & 0 & 0.43 & 0 & 0 & 0.4 & 0.75 & 1 & 0.5 & 1 \\ 0.83 & 0.2 & 0.4 & 0.33 & 0.25 & 0.4 & 0.43 & 0.71 & 0.8 & 0.6 & 0 & 0.75 & 1 & 0.5 & 1 \\ 0.67 & 0.6 & 0.2 & 0.67 & 0.5 & 0 & 0.14 & 0.29 & 0 & 0.8 & 0.6 & 0.75 & 1 & 0.5 & 1 \end{bmatrix}^T$$

$$B_3 = \begin{bmatrix} 0.67 & 0.8 & 0.2 & 0 & 1 & 0.4 & 0.71 & 0.29 & 0 & 1 & 0.4 & 1 & 0.5 & 0 & 0 \\ 0.33 & 0.4 & 0 & 0.33 & 0.88 & 0.6 & 0.57 & 0.57 & 0.4 & 0.8 & 0.8 & 0.1 & 0.5 & 0 & 0 \\ 0.5 & 0.6 & 0.4 & 0.67 & 0.75 & 0.8 & 0.86 & 0.71 & 0.6 & 0.4 & 1 & 0 & 0.5 & 0 & 0 \\ 0.83 & 0.2 & 0.6 & 0.83 & 0.63 & 0.2 & 1 & 0.86 & 0.2 & 0.2 & 0.2 & 1 & 0.5 & 0 & 0 \\ 1 & 1 & 0.8 & 0.5 & 0.5 & 0.6 & 0.71 & 1 & 1 & 0.4 & 0.2 & 1 & 0.5 & 0 & 0 \\ 0.83 & 0.4 & 0.4 & 0.71 & 0.83 & 1 & 0.29 & 0.43 & 0.2 & 0.6 & 0.4 & 1 & 0.5 & 0 & 0 \end{bmatrix}^T$$

From expression (5), the weighted standardized evaluation matrices are derived:

$$C_1 = \begin{bmatrix} 0 & 0 & 0.01 & 0.01 & 0 & 0 & 0.01 & 0 & 0.01 & 0 & 0.01 & 0.01 & 0 & 0.01 & 0.01 \\ 0.01 & 0.01 & 0.01 & 0.01 & 0 & 0.01 & 0 & 0 & 0.01 & 0.01 & 0 & 0.01 & 0 & 0.01 & 0 \\ 0.01 & 0.01 & 0.01 & 0.01 & 0.01 & 0.01 & 0.01 & 0.01 & 0.01 & 0.01 & 0.01 & 0.01 & 0 & 0.01 & 0.01 \\ 0.01 & 0.01 & 0.01 & 0.01 & 0 & 0 & 0 & 0 & 0.01 & 0.01 & 0 & 0 & 0 & 0 & 0 \\ 0.02 & 0 & 0.02 & 0.01 & 0 & 0.01 & 0 & 0.01 & 0.01 & 0 & 0 & 0.01 & 0 & 0.01 & 0 \\ 0.01 & 0 & 0 & 0 & 0 & 0 & 0 & 0 & 0 & 0 & 0 & 0 & 0 & 0 & 0 \end{bmatrix}^T$$

$$C_1 = \begin{bmatrix} 0.01 & 0.01 & 0.01 & 0.01 & 0 & 0 & 0 & 0 & 0 & 0.01 & 0 & 0.01 & 0.01 & 0 & 0.01 \\ 0.01 & 0 & 0.01 & 0.01 & 0 & 0 & 0 & 0 & 0 & 0 & 0.01 & 0.01 & 0.01 & 0 & 0.01 \\ 0.01 & 0.01 & 0 & 0.01 & 0.01 & 0.01 & 0.01 & 0 & 0 & 0.01 & 0.01 & 0.01 & 0.01 & 0.01 & 0.01 \\ 0.01 & 0.01 & 0 & 0 & 0 & 0 & 0 & 0 & 0 & 0 & 0 & 0 & 0.01 & 0 & 0 \\ 0.02 & 0 & 0.01 & 0 & 0 & 0 & 0 & 0.01 & 0.01 & 0 & 0 & 0.01 & 0.01 & 0 & 0.01 \\ 0.01 & 0.01 & 0 & 0 & 0 & 0 & 0 & 0 & 0 & 0 & 0 & 0 & 0 & 0 & 0 \end{bmatrix}^T$$

$$C_3 = \begin{bmatrix} 0.02 & 0.02 & 0 & 0 & 0.01 & 0 & 0.01 & 0 & 0 & 0.01 & 0 & 0.01 & 0 & 0 & 0 \\ 0.01 & 0.01 & 0 & 0 & 0.01 & 0.01 & 0 & 0.01 & 0 & 0.01 & 0 & 0 & 0 & 0 & 0 \\ 0.02 & 0.02 & 0.01 & 0.01 & 0.01 & 0.01 & 0.01 & 0.01 & 0.01 & 0 & 0.01 & 0 & 0.01 & 0 & 0 \\ 0.01 & 0 & 0.01 & 0.01 & 0 & 0 & 0.01 & 0.01 & 0 & 0 & 0 & 0.01 & 0 & 0 & 0 \\ 0.03 & 0.02 & 0.02 & 0.01 & 0.01 & 0.01 & 0.01 & 0.01 & 0.01 & 0 & 0 & 0.01 & 0 & 0 & 0 \\ 0.01 & 0 & 0 & 0 & 0 & 0.01 & 0 & 0 & 0 & 0 & 0 & 0 & 0 & 0 & 0 \end{bmatrix}^T$$

From expression (6), the positive ideal matrix $C^+$ and negative ideal matrix $C^-$ are determined as follows:

$$C^+ = \begin{bmatrix} 0.02 & 0.02 & 0.01 & 0.01 & 0.01 & 0 & 0.01 & 0 & 0.01 & 0.01 & 0.01 & 0.01 & 0.01 & 0.01 & 0.01 \\ 0.01 & 0.01 & 0.01 & 0.01 & 0.01 & 0.01 & 0 & 0.01 & 0.01 & 0.01 & 0.01 & 0.01 & 0.01 & 0.01 & 0.01 \\ 0.02 & 0.02 & 0.01 & 0.01 & 0.01 & 0.01 & 0.01 & 0.01 & 0.01 & 0.01 & 0.01 & 0.01 & 0.01 & 0.01 & 0.01 \\ 0.01 & 0.01 & 0.01 & 0.01 & 0 & 0.01 & 0.01 & 0.01 & 0.01 & 0.01 & 0 & 0.01 & 0.01 & 0.01 & 0.01 \\ 0.03 & 0.02 & 0.02 & 0.01 & 0.01 & 0.01 & 0.01 & 0.01 & 0.01 & 0 & 0 & 0.01 & 0.01 & 0.01 & 0.01 \\ 0.01 & 0.01 & 0 & 0 & 0 & 0.01 & 0 & 0 & 0 & 0 & 0 & 0 & 0 & 0 & 0 \end{bmatrix}^T$$

$$C^- = \begin{bmatrix} 0 & 0 & 0 & 0 & 0 & 0 & 0 & 0 & 0 & 0 & 0 & 0.01 & 0 & 0 & 0 \\ 0.01 & 0 & 0 & 0 & 0 & 0 & 0 & 0 & 0 & 0 & 0 & 0 & 0 & 0 & 0 \\ 0.01 & 0.01 & 0 & 0.01 & 0.01 & 0.01 & 0.01 & 0 & 0 & 0 & 0.01 & 0 & 0 & 0 & 0 \\ 0.01 & 0 & 0 & 0 & 0 & 0 & 0 & 0 & 0 & 0 & 0 & 0 & 0 & 0 & 0 \\ 0.02 & 0 & 0.01 & 0 & 0 & 0 & 0 & 0.01 & 0.01 & 0 & 0 & 0.01 & 0 & 0 & 0 \\ 0.01 & 0 & 0 & 0 & 0 & 0 & 0 & 0 & 0 & 0 & 0 & 0 & 0 & 0 & 0 \end{bmatrix}^T$$

From expression (7), the local volume matrices corresponding to each weighted standardized evaluation matrix and the positive/negative ideal matrices are presented below.

$$D_1 = \begin{bmatrix} 0 & 0 & 0 & 0 & 0 & 0 & 0 & 0 & 0 & 0 & 0 & 0 & 0 & 0 \\ 0.01 & 0 & 0 & 0 & 0 & 0 & 0 & 0 & 0 & 0 & 0 & 0 & 0 & 0 \\ 0.01 & 0 & 0 & 0 & 0 & 0 & 0 & 0 & 0 & 0 & 0 & 0 & 0 & 0 \\ 0.01 & 0 & 0 & 0 & 0 & 0 & 0 & 0 & 0 & 0 & 0 & 0 & 0 & 0 \\ 0.01 & 0 & 0 & 0 & 0 & 0 & 0 & 0 & 0 & 0 & 0 & 0 & 0 & 0 \end{bmatrix}^T$$

$$D_2 = \begin{bmatrix} 0 & 0 & 0 & 0 & 0 & 0 & 0 & 0 & 0 & 0 & 0 & 0 & 0 & 0 \\ 0 & 0 & 0 & 0 & 0 & 0 & 0 & 0 & 0 & 0 & 0 & 0 & 0 & 0 \\ 0.01 & 0 & 0 & 0 & 0 & 0 & 0 & 0 & 0 & 0 & 0 & 0 & 0 & 0 \\ 0.01 & 0 & 0 & 0 & 0 & 0 & 0 & 0 & 0 & 0 & 0 & 0 & 0 & 0 \\ 0 & 0 & 0 & 0 & 0 & 0 & 0 & 0 & 0 & 0 & 0 & 0 & 0 & 0 \end{bmatrix}^T$$

$$D_3 = \begin{bmatrix} -0.01 & -0.01 & 0 & 0 & 0 & 0 & 0 & 0 & 0 & 0 & 0 & 0 & 0 & 0 \\ -0.01 & 0 & 0 & 0 & 0 & 0 & 0 & 0.01 & 0 & 0 & 0 & 0 & 0 & 0 \\ -0.01 & 0 & 0.01 & 0 & 0 & 0 & 0 & 0.01 & 0 & 0 & 0 & 0 & 0 & 0 \\ -0.01 & 0 & 0.01 & 0 & 0 & 0 & 0 & 0.01 & 0 & 0 & 0 & 0 & 0 & 0 \\ 0 & 0 & 0 & 0 & 0 & 0 & 0 & 0 & 0 & 0 & 0 & 0 & 0 & 0 \end{bmatrix}^T$$

$$D^+ = \begin{bmatrix} -0.01 & 0 & 0 & 0 & 0 & 0 & 0 & 0 & 0 & 0 & 0 & 0 & 0 & 0 \\ -0.01 & 0 & 0 & 0 & 0 & 0 & 0 & 0 & 0 & 0 & 0 & 0 & 0 & 0 \\ -0.01 & -0.01 & 0 & 0 & 0 & 0 & 0 & 0 & 0 & 0 & 0 & 0 & 0 & 0 \\ 0 & 0 & 0 & 0 & 0 & 0 & 0 & 0 & 0 & 0 & 0 & 0 & 0 & 0 \\ 0 & -0.01 & 0 & -0.01 & 0 & 0 & 0 & 0 & 0 & 0 & 0 & 0 & 0 & 0 \end{bmatrix}^T$$

$$D^- = \begin{bmatrix} 0 & 0 & 0 & 0 & 0 & 0 & 0 & 0 & 0 & 0 & 0 & 0 & 0 & 0 \\ 0 & 0 & 0 & 0 & 0 & 0 & 0 & 0 & 0 & 0 & 0 & 0 & 0 & 0 \\ 0.01 & 0 & 0 & 0 & 0 & 0 & 0 & 0 & 0 & 0 & 0 & 0 & 0 & 0 \\ 0.01 & 0 & 0 & 0 & 0 & 0 & 0 & 0 & 0 & 0 & 0 & 0 & 0 & 0 \\ 0.01 & 0 & 0 & 0 & 0 & 0 & 0 & 0 & 0 & 0 & 0 & 0 & 0 & 0 \end{bmatrix}^T$$

From expression (8), the volume difference matrices between the local volume matrices of each weighted standardized evaluation matrix and the local volume matrix of the positive ideal matrix can be derived as follows.

$$D^+_{(1)} = \begin{bmatrix} 0.01 & 0.01 & 0 & 0 & 0 & 0 & 0 & 0 & 0 & 0 & 0 & 0 & 0 & 0 \\ 0.01 & 0.01 & 0 & 0 & 0 & 0 & 0 & 0 & 0 & 0 & 0 & 0 & 0 & 0 \\ 0.01 & 0.01 & 0 & 0 & 0 & 0 & 0 & 0 & 0 & 0 & 0 & 0 & 0 & 0 \\ 0.01 & 0 & 0 & 0 & 0 & 0 & 0 & 0 & 0 & 0 & 0 & 0 & 0 & 0 \\ 0.01 & 0.01 & 0 & 0 & 0 & 0 & 0 & 0 & 0 & 0 & 0 & 0 & 0 & 0 \end{bmatrix}^T$$

$$D^+_{(2)} = \begin{bmatrix} 0.01 & 0 & 0 & 0 & 0 & 0 & 0 & 0 & 0 & 0 & 0 & 0 & 0 & 0 \\ 0.01 & 0 & 0 & 0 & 0 & 0 & 0 & 0 & 0 & 0 & 0 & 0 & 0 & 0 \\ 0.01 & 0 & 0 & 0 & 0 & 0 & 0 & 0 & 0 & 0 & 0 & 0 & 0 & 0 \\ 0.01 & 0 & 0.01 & 0 & 0 & 0 & 0 & 0 & 0 & 0 & 0 & 0 & 0 & 0 \\ 0.01 & 0 & 0 & 0 & 0 & 0 & 0 & 0 & 0 & 0 & 0 & 0 & 0 & 0 \end{bmatrix}^T$$

$$D^+_{(3)} = \begin{bmatrix} 0 & 0 & 0 & 0 & 0 & 0 & 0 & 0 & 0 & 0 & 0 & 0 & 0 & 0 \\ 0 & 0 & 0 & 0 & 0 & 0 & 0 & 0 & 0 & 0 & 0 & 0 & 0 & 0 \\ 0 & 0 & 0.01 & 0 & 0 & 0 & 0 & 0 & 0 & 0 & 0 & 0 & 0 & 0 \\ 0 & 0 & 0.01 & 0 & 0 & 0 & 0 & 0 & 0 & 0 & 0 & 0 & 0 & 0 \\ 0 & 0 & 0.01 & 0 & 0 & 0 & 0 & 0 & 0 & 0 & 0 & 0 & 0 & 0 \end{bmatrix}^T$$

From expression (9), the grey incidence matrices between each weighted standardized evaluation matrix and the positive ideal matrix can be obtained as follows.

$$G^+_{(1)} = \begin{bmatrix} 0.35 & 0.44 & 0.89 & 0.93 & 0.86 & 0.96 & 0.91 & 0.97 & 0.93 & 0.89 & 0.96 & 0.94 & 0.94 & 0.99 \\ 0 & 0.44 & 0.97 & 0.91 & 0.83 & 0.96 & 0.91 & 0.96 & 0.89 & 0.91 & 0.99 & 0.92 & 0.99 & 0.99 \\ 0.21 & 0.43 & 0.95 & 0.81 & 0.82 & 0.94 & 0.92 & 0.89 & 0.89 & 0.83 & 0.97 & 0.97 & 0.93 & 1.00 \\ 0.17 & 0.65 & 0.87 & 0.68 & 0.68 & 0.93 & 0.92 & 0.99 & 0.82 & 0.84 & 0.99 & 0.89 & 0.92 & 0.97 \\ 0.30 & 0.43 & 0.92 & 0.72 & 0.69 & 0.89 & 0.89 & 0.93 & 0.89 & 0.90 & 0.99 & 0.89 & 0.88 & 0.98 \end{bmatrix}^T$$

$$G^+_{(2)} = \begin{bmatrix} 0.54 & 0.88 & 0.85 & 0.92 & 0.86 & 0.94 & 0.90 & 0.98 & 0.97 & 0.96 & 0.95 & 0.96 & 0.99 & 0.97 \\ 0.29 & 0.95 & 0.71 & 0.75 & 0.80 & 0.90 & 0.98 & 0.87 & 1.00 & 0.89 & 0.99 & 0.96 & 0.98 & 1.00 \\ 0.25 & 0.89 & 0.67 & 0.76 & 0.76 & 0.98 & 0.84 & 0.80 & 0.96 & 0.97 & 0.95 & 0.97 & 1.00 & 0.97 \\ 0.28 & 0.79 & 0.62 & 0.86 & 0.69 & 0.99 & 0.98 & 0.89 & 0.98 & 0.91 & 0.98 & 0.98 & 0.95 & 0.96 \\ 0.43 & 0.86 & 0.76 & 0.68 & 0.66 & 0.96 & 0.90 & 0.97 & 0.83 & 0.88 & 0.95 & 0.96 & 0.94 & 0.89 \end{bmatrix}^T$$

$$G^+_{(3)} = \begin{bmatrix} 0.92 & 0.95 & 0.96 & 0.90 & 1.00 & 1.00 & 1.00 & 0.96 & 0.92 & 0.98 & 0.99 & 0.89 & 0.94 & 0.92 \\ 0.96 & 0.98 & 0.65 & 0.80 & 1.00 & 1.00 & 1.00 & 0.84 & 0.85 & 0.98 & 0.86 & 0.73 & 0.97 & 0.97 \\ 0.93 & 0.88 & 0.54 & 0.71 & 1.00 & 1.00 & 1.00 & 0.81 & 0.99 & 0.95 & 0.86 & 0.89 & 0.96 & 0.93 \\ 0.85 & 0.87 & 0.48 & 0.81 & 1.00 & 1.00 & 1.00 & 0.85 & 0.93 & 0.99 & 0.93 & 0.95 & 0.86 & 0.83 \\ 0.97 & 0.73 & 0.52 & 0.82 & 0.98 & 1.00 & 1.00 & 0.76 & 0.81 & 0.97 & 0.92 & 0.95 & 0.82 & 0.75 \end{bmatrix}^T$$

From expression (10), the volumetric incidence degree between each weighted standardized evaluation matrix and the positive ideal matrix can be derived as follows.

$$\gamma^+_{01}=0.89, \quad \gamma^+_{02}=0.92, \quad \gamma^+_{03}=0.96$$

Similarly, the volumetric incidence degree between each weighted standardized evaluation matrix and the negative ideal matrix can be obtained as follows.

$$\gamma^-_{01}=0.97, \quad \gamma^-_{02}=0.93, \quad \gamma^-_{03}=0.89$$

From expression (11), the superiority degree of each weighted standardized evaluation matrix relative to the positive ideal matrix can be derived as follows.

$$s_1=0.46, \quad s_2=0.50, \quad s_3=0.54$$

Thus, among 3 evaluated areas, area 3 exhibits the highest risk level, followed by area 2, while area 1 demonstrates the lowest risk. According to the risk classification principle outlined in Section 4.4, the risk levels of 3 evaluated areas are categorized as medium.

## 6. Conclusion

This paper addresses the challenge of dynamic fire risk assessment in the WUI. By applying the ideal matrix method, a matrix similarity incidence analysis model for dynamic fire risk assessment is developed. The fire risk of each evaluated area is ranked based on the incidence coefficients between their weighted standardized evaluation matrices and both the positive and negative ideal matrices, with corresponding risk levels assigned. The results indicate that the matrix similarity incidence analysis model not only simplifies the computational process but also effectively captures the dynamic evolution of fire risks across different evaluated areas.

## Conflict of Interests

The authors declare that there is no conflict of interests regarding the publication of this paper.

# Acknowledgments

This work was supported by National Natural Science Foundation of China (No. 72174031), the Fundamental Research Funds for the Central Universities (No. 2023ZKPYLX02).